\documentclass[12pt,a4paper]{article}
\pdfoutput=1
\usepackage{fixltx2e}
\usepackage{float}
\usepackage{jheppub}
\usepackage{multirow}
\usepackage{chemarrow}
\usepackage{amsmath,amssymb,amsfonts,mathtools}
\usepackage{subfig}
\usepackage{scalefnt}
\usepackage{wrapfig}
\usepackage{fancybox}
\usepackage{ifsym}
\newcommand{\bea}{\begin{eqnarray}\displaystyle}
\newcommand{\eea}{\end{eqnarray}}

\newlength{\arrow}
{\setlength{\fboxsep}{15pt}
\setlength{\mylength}{\linewidth}%
\addtolength{\mylength}{-2\fboxsep}%
\addtolength{\mylength}{-2\fboxrule}%
\Sbox
\minipage{\mylength}%
\setlength{\abovedisplayskip}{0pt}%
\setlength{\belowdisplayskip}{0pt}%
\equation}%
{\endequation\endminipage\endSbox
\[\fbox{\TheSbox}\]}

\usepackage{epsfig}
\usepackage{tikz}
\usetikzlibrary{decorations.pathmorphing}

\usepackage{float}

\def\2{{1\over2}}
\let\<=\langle \let\>=\rangle
\def\new#1\endnew{{\bf #1}}
\def\ifundefined#1{\expandafter\ifx\csname#1\endcsname\relax}
\ifundefined{draftmode}\else    \input draftmode        \fi
\let\Msize=\footnotesize             
\def\BM{\Msize\begin{matrix}}           \def\EM{\end{matrix}}
\def\MN M:#1 #2 N:#3 #4 {{(#1_{#2},#3_{#4})}}
\def\MNH M:#1 #2 N:#3 #4 H:#5,#6 [#7]{{(#1_{#2},#3_{#4})^{#5,#6}_{#7}}}

\def\CN{{\cal N}}

\newcommand{\be}{\begin{equation}}
\newcommand{\ee}{\end{equation}}

\title{From strings in 6d to strings in 5d}
\author[\ast\dagger]{Babak Haghighat,}
\affiliation[\ast]{Jefferson Physical Laboratory, Harvard University, Cambridge, MA 02138, USA}
\affiliation[\dagger]{Department of Mathematics, Harvard University, Cambridge, MA 02138, USA}

\abstract{We show how recent progress in computing elliptic genera of strings in six dimensions can be used to obtain expressions for elliptic genera of strings in five-dimensional field theories which have a six-dimensional parent. We further connect our results to recent mathematical results about sheaf counting on ruled surfaces.
}

\begin{document}
\maketitle

%\tableofcontents

\section{Introduction}

Strings of 5d supersymmetric Yang-Mills theories have been studied in the past to some extend \cite{Seiberg:1996bd,Boyarsky:2002ck,Lambert:2010iw,Haghighat:2011xx,Haghighat:2012bm}. Among the main important questions regarding these strings are whether they form bound states. In particular one is interested in bound states of BPS strings which become particles of finite mass in 4d once the five-dimensional theory is compactified on a circle. %As the strings carry magnetic charge which becomes monopole charge in four dimensions their bound states at zero momentum along the circle is determined by the BPS spectrum of the four-dimensional gauge theory.%
 One of the interesting questions is whether the strings only form bound states once they are given momentum along the circle. Upon further compactification of the five-dimensional theory on a two-torus $r$ strings can wrap the torus and contribute to the metric on the Coulomb branch of the resulting three-dimensional theory as instanton corrections \cite{Haghighat:2011xx} weighted by the exponent of minus the instanton action:
\begin{equation}
	e^{-\mathcal{V}r T - i r \lambda},
\end{equation}
where $\mathcal{V}$ is the volume of the torus, $T$ is the tension of the strings and $\lambda$ the dual to the three-dimensional vector. Furthermore, note that the above exponential comes with a pre-factor given by the elliptic genus $Z_r$ of $r$ strings wrapping the torus. Computing these $Z_r$ is in general a very complicated task as one has to compute the elliptic genus of a sigma model whose target space is the moduli space of magnetic monopoles \cite{Haghighat:2011xx,Haghighat:2012bm}. 

However, there is an alternative to this whenever one knows how to geometrically engineer the five-dimensional quantum field theory. Within the framework of geometric engineering the strings are realized as M5 branes wrapping a four-cycle in a non-compact Calabi-Yau threefold and the elliptic genus of the strings gets related to the partition function of twisted $\mathcal{N}=4$ SYM on the four-manifold \cite{Minahan:1998vr}. Such partition functions are mathematically generating functions of sheaves on algebraic surfaces, and recently there has been a lot of progress in obtaining them \cite{Manschot:2010nc,Manschot:2011dj,Manschot:2011ym,Manschot:2014cca}. Such sheaf counting can then be applied to surfaces which are relevant for geometrically engineering the five-dimensional theory in order to obtain the elliptic genus of the magnetic string as has been done in \cite{Haghighat:2012bm} for the case of del Pezzo surfaces.

In this paper we want to put forward yet another prescription for obtaining the elliptic genera $Z_r$ for a subclass of five-dimensional theories which can be obtained from a six-dimensional parent through circle reduction. The six-dimensional parent theory has itself strings and in a compactification to five dimensions they may or may not wrap the circle. In case they do wrap the circle their elliptic genus can be computed through various techniques which have recently been developed \cite{ Haghighat:2013gba,Haghighat:2013tka,Haghighat:2014pva,Haghighat:2014vxa} and allow to obtain the partition function of the six-dimensional theory on $T^2 \times \mathbb{R}^4$:
\begin{equation}
	Z^{6d}_{\mathbb{R}^4\times T^2} = \sum_n e^{-n \phi} Z^n_{T^2},
\end{equation}
where $Z^n_{T^2}$ denotes the elliptic genus of $n$ strings and $\phi$ is the vev of the scalar in the 6d tensor multiplet. In case the strings do not wrap the circle from 6d to 5d they become strings of the five-dimensional theory. Our claim is that the elliptic genus of the strings in 5d can be obtained from the elliptic genus of the strings in 6d in the Nekrasov-Shatashvilli limit and we will demonstrate this explicitly for the case of the 5d $\mathcal{N}=1^*$ $SU(2)$ gauge theory. In doing this we connect to recent mathematical results on generating functions of Poincar\'e polynomials of moduli spaces of sheaves on ruled surfaces \cite{Mozgovoy:2013zqx}. Perturbative expansions of elliptic genera for strings of the 5d $\mathcal{N}=1^*$ theory have been obtained in the past using instanton calculus \cite{Kim:2011mv}, but our results are more complete since they provide analytic expressions for elliptic genera of any magnetic charge.

The organization of this paper is as follows: In Section 2 we explain the duality which connects the strings in 6d to the strings in 5d. In Section 3 we derive expressions for the corresponding elliptic genera and finally in Section 4 we elaborate on a specific example.

\section{A 6d/5d duality}
\label{sec:6d5d}

In this section we want to review the geometric engineering of five-dimensional quantum field theories which have a six-dimensional parent and the duality which connects the two theories in 6d and 5d. The duality we want to discuss is a special case of the duality considered in \cite{Witten:1996bn} where the Calabi-Yau fourfold is taken to be $CY_3 \times T^2$.

Consider M-theory compactified on a non-compact elliptic Calabi-Yau manifold $X$ with a section $\pi: X \rightarrow B$ and with fibers given by two-tori $E$. Here $B$ is a non-compact complex two-fold of the form 
\begin{equation}
	B = \mathcal{O}(-n) \rightarrow \mathbb{P}^1,
\end{equation}
with $n$ a positive integer. Alternatively, $X$ can be viewed as the anti-canonical bundle over a surface $D$ which is locally given by the product of the elliptic fiber $E$ and the $\mathbb{P}^1$. This setup leads to a five-dimensional quantum field theory with eight supercharges, that is with $\mathcal{N} = 1$ supersymmetry. Furthermore, the five-dimensional field theory has BPS strings \cite{Seiberg:1996bd} which are M5 branes wrapped around $D$. Their Central charge is given by their number $r$ times their tension $T$:
\begin{equation}
	Z_{strings} = r T = r \textrm{Vol}(D).
\end{equation}
Consider for example the case $n=2$ which is also the main example of this paper. This case is special in that it gives rise to a five-dimensional theory with $\mathcal{N} = 2$ supersymmetry, namely 5d $\mathcal{N}=2$ $SU(2)$ gauge theory. However, as will be reviewed below, there is a ``mass-deformation" which leads to the $\mathcal{N}=1^*$ theory. The strings of this theory are uplifts of the four-dimensional magnetic monopole solution.

Next, we want to follow a chain of dualities which connect M-theory on $X$ with Type IIB on $B$. We start by applying the fiber-wise duality between M-theory on $E$ and Type IIB on $S^1$: Take the elliptic curve $E$ to be a rectangular torus
\begin{equation}
	E = S^1_{\tilde R_1} \times S^1_{\tilde R_2},
\end{equation}
and compactify on $S^1_{\tilde R_1}$ to Type IIA. Then the Type IIA coupling constant will be (setting $\alpha' = 1$) $\lambda_{IIA} = \tilde R_1$. T-dualizing along $S^1_{\tilde R_2}$ one arrives at Type IIB on $\mathbb{R}^5 \times S^1_{\frac{1}{\tilde R_2}} \times B$ with coupling $\lambda_{IIB} = \lambda_{IIA}/\tilde R_2$. In fact $\lambda_{IIB}$ varies with the position on $B$ and the resulting vacuum can be seen as compactifying F-theory first on $X$, as studied in \cite{Morrison:2012np}, and then further on $S^1_{\frac{1}{\tilde R_2}}$. We are particularly interested in the fate of the M5 branes which give rise to strings in 5d along this chain of dualities. These become D4 branes upon compactification to Type IIA and then subsequently D3 branes wrapping $\mathbb{P}^1$ after T-dualizing to Type IIB. As D3 branes wrapping $\mathbb{P}^1$ give rise to strings of minimal 6d SCFTs which have recently been studied in \cite{Haghighat:2014vxa}, we have managed to map strings in 5d to strings in 6d. To summarize we arrive at the following duality:
\begin{table}[h]
\begin{center}
\begin{tabular}{cccc}
    ~                       & 6d                                         & ~ & 5d \\
     \vspace{2mm}
    ~                       & Type IIB on $B \times S^1\times \mathbb{R}^5$   & ~ & M-theory on $X\times \mathbb{R}^4$ \\
   strings:              &  D3/$\mathbb{P}^1$    & $\longleftrightarrow$ &     M5/$D$ \\
\end{tabular}
\end{center}
%\caption{Relation between strings in 6d and 5d.}
\end{table}

\subsection*{Example: 5d $\mathcal{N}=1^*$ $SU(2)$ gauge theory}

Let us now present an example to illuminate the idea presented above. Consider the case $n=2$, that is take $B = \mathcal{O}(-2) \rightarrow \mathbb{P}^1$ to be the resolved $A_1$ singularity. Type IIB compactified on this manifold gives rise to the $(2,0)$ superconformal theory and a further compactification on a circle gives maximal Super-Yang-Mills in five dimensions. The dual M-theory has to be compactified on $E \times B$. The mass deformation to the $\mathcal{N}=1^*$ theory can be introduced by making $E$ singular of Kodaira type $I_1$ along $\mathbb{P}^1 \subset B$. The K\"ahler class of the resolution of the $I_1$ fiber corresponds in this picture to the mass $m$ of the adjoint hypermultiplet. The relevant Calabi-Yau has been discussed in the language of toric geometry in \cite{Haghighat:2013gba}. The underlying toric diagram there is depicted in Figure 1 (a). 

\begin{figure}[h]
  \centering
  \subfloat[The toric diagram]{\label{fig:toric}\includegraphics[width=0.5\textwidth]{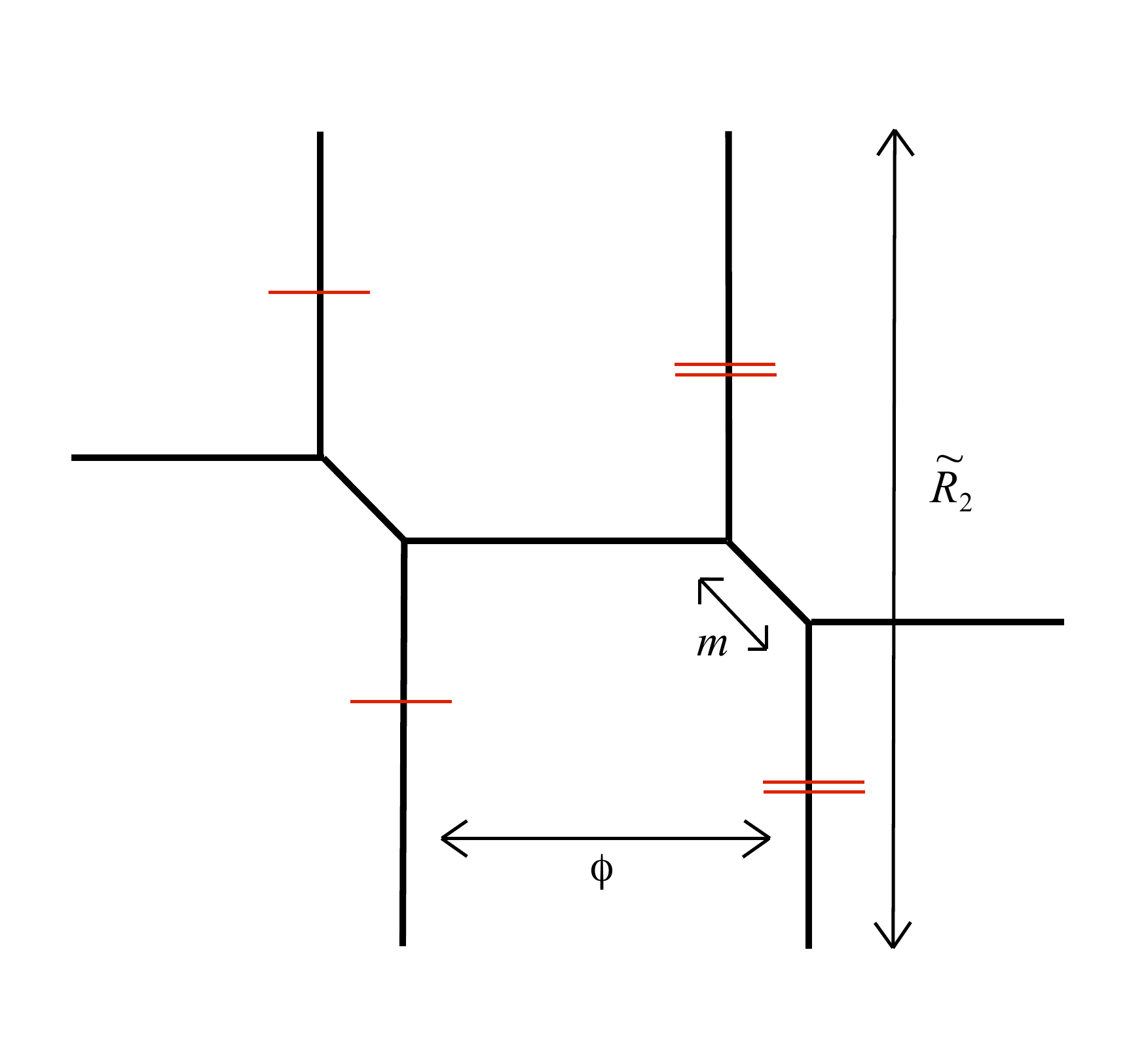}}     
  %\hspace{.1in}    
  \subfloat[The M5 brane locus]{\label{fig:M5}\includegraphics[width=0.5\textwidth]{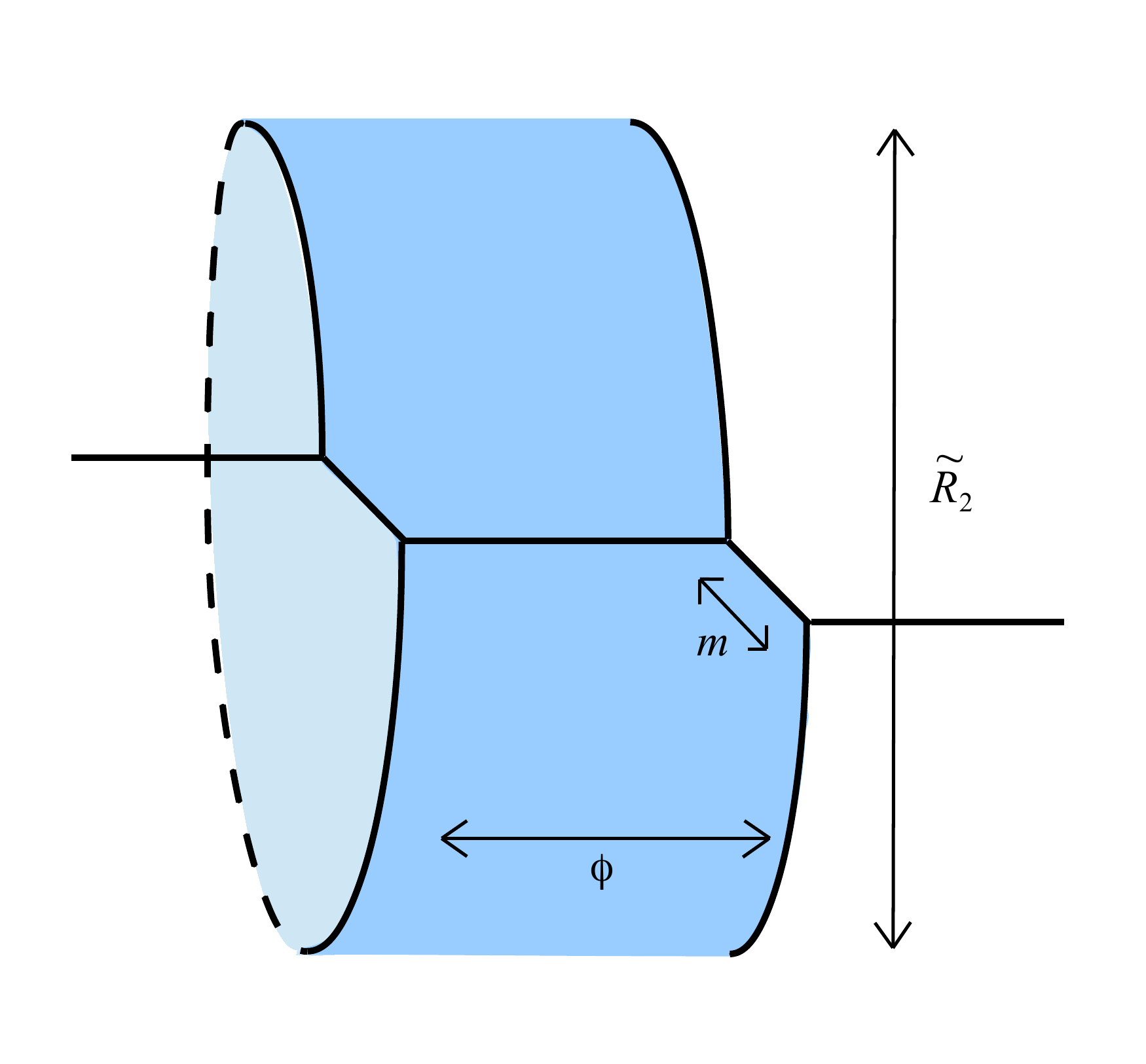}}
  \caption{In (a) we see the toric diagram for the Calabi-Yau which geometrically engineers 5d $\mathcal{N}=1^*$ $SU(2)$ gauge theory. The red marks denote that the corresponding legs are identified. In (b) we see the same toric geometry but this time the locus of the M5 brane wrapping $D = E \times \mathbb{P}^1$ is marked blue.}
  \label{fig:Mstrings}
\end{figure}

The edges of the toric skeleton show the locus where a circle of a two-torus fiber degenerates and the vertices are points where the whole two-torus degenerates. Therefore the circular direction with length $\tilde R_2$ actually describes the elliptic fiber $E$. In our toric description $E$ is a circle fibration over this circular direction with total volume $\tilde R_1 \tilde R_2$ which is identified in the geometric engineering picture with the Yang-Mills coupling, i.e. $\tilde R_1 \tilde R_2 = \frac{1}{g_{YM}^2}$. Furthermore, strings in 5d arise by wrapping an M5 brane on the four-cycle depicted in blue in Figure 1 (b) and will therefore have the following tension:
\begin{equation}
	T = \textrm{vol}(E\times\mathbb{P}^1) = \frac{\phi}{g_{YM}^2}.
\end{equation}

\section{Elliptic genus of the 5d string}

In the previous section we saw that the strings in 6d get related to the strings in 5d. In this section we want to study their elliptic genus by further compactifying both theories on a two torus $T^2$. The elliptic genera will then contribute to the partition function/moduli space of the resulting four- and three-dimensional theories. We start by taking the radii of the torus to be as follows where for simplicity we focus on the rectangular case
\begin{equation}
	T^2 = S^1_{R_1} \times S^1_{R_2}, 
\end{equation}
and identify the complex structure of $T^2$ with the volume of $E$, which is a constant, giving\footnote{Normalizing $R_1=\tilde R_1 = 1$ the relation follows from the F-theory/M-theory duality relation $\frac{1}{R_2}=\tilde R_2$ as discussed in the previous section.} 
\begin{equation}
	\tau = i \frac{R_1}{R_2} = i \tilde R_1 \tilde R_2.
\end{equation}
The resulting theories live now on $T^2 \times \mathbb{R}^4$ and $T^2 \times \mathbb{R}^3$. We can now introduce a twisting of $\mathbb{R}^4 \cong \mathbb{C} \times \mathbb{C}$ (with coordinates $z_1$ and $z_2$) along the cycles of $T^2$ as done in \cite{Haghighat:2013gba}:
\begin{equation}
	U(1)_{\epsilon_1}\times U(1)_{\epsilon_2} : (z_1,z_2) \mapsto(e^{2\pi i \epsilon_1} z_1,e^{2\pi i \epsilon_2} z_2),
\end{equation}
and similarly for $\mathbb{R}^3 \cong \mathbb{C} \times \mathbb{R}$ (with coordinates $z_2$ and $x$) where only one of the $U(1)_{\epsilon_i}$ can act:
\begin{equation}
	U(1)_{\epsilon_2} : (x,z_2) \mapsto (x, e^{2\pi i \epsilon_2} z_2).
\end{equation}
The situation for the two cases is shown below in Figure 2.
\begin{figure}[h]
  \centering
  \subfloat[6d]{\label{fig:toric}\includegraphics[width=0.45\textwidth]{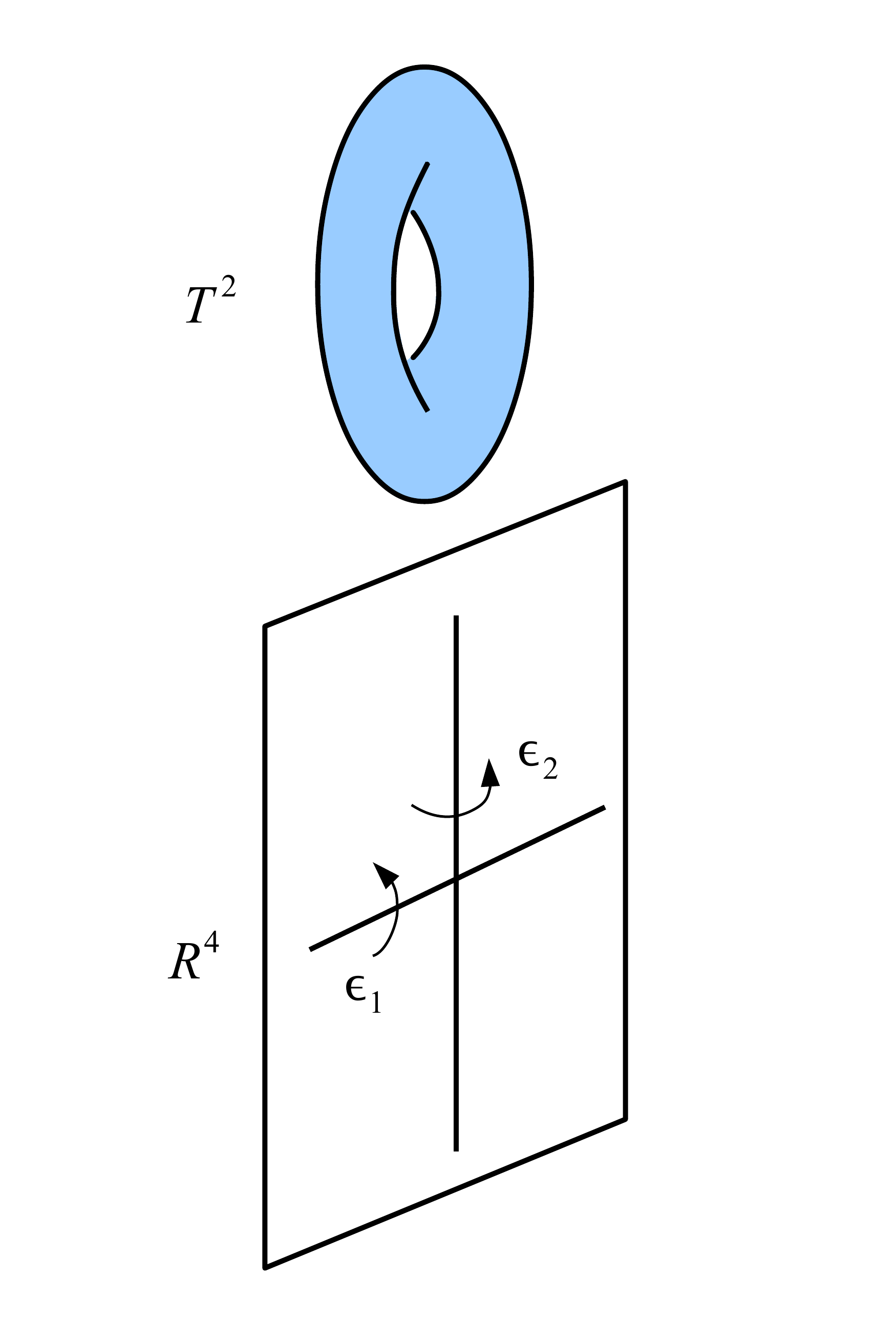}}     
  \hspace{.1in}    
  \subfloat[5d]{\label{fig:M5}\includegraphics[width=0.45\textwidth]{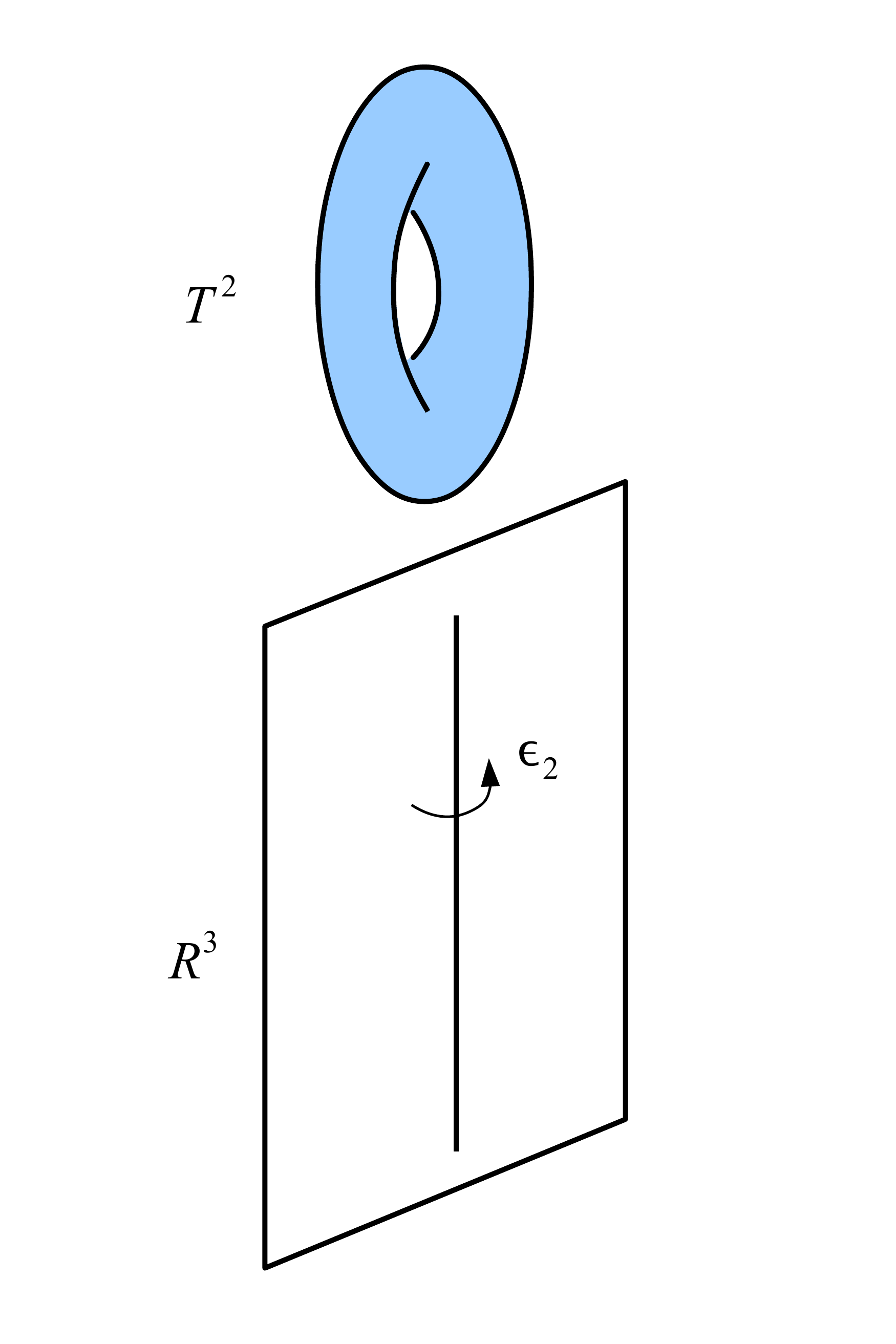}}
  \caption{In (a) we see the strings of the six-dimensional theory wrapping a $T^2$ and probing $\mathbb{R}^4$. In (b) we see strings of the five-dimensional theory wrapping the two-torus and probing $\mathbb{R}^3$.}
  \label{fig:6d5d}
\end{figure}
The picture suggests a very simple relation between the two elliptic genera. Recall that the elliptic genus, which is a supersymmetric protected index, of the strings in 6d is defined as \cite{Haghighat:2013gba}:
\begin{equation} \label{eq:6dgenus}
	Z_n(Q_{\tau},\epsilon_1,\epsilon_2,m_l) = \textrm{Tr}_{\textrm{RR}}\left[(-1)^F Q_{\tau}^{H_L} \overline{Q}_{\tau}^{H_R} e^{2\pi i \epsilon_1 J_1} e^{2\pi i \epsilon_2 J_2} y^{J_I} \prod_{l=1}^N e^{2\pi i m_l F_l}\right],
\end{equation}
where $Q_{\tau} = e^{2\pi i \tau}$, $J_1$, $J_2$ are the Cartans of $SO(4)_{\mathbb{R}^4}=SU(2)_1 \times SU(2)_2$, $J_I$ is the Cartan of an $SU(2)$ R-symmetry\footnote{The dependence on $y$ may or may not be present. For example, the E-string theory does not couple to the $J_I$-current. On the other and, in the case of the M-string $y$ can be identified with the mass-deformation parameter $Q_m$ as explained in \cite{Haghighat:2013gba}.} with fugacity $y=e^{2\pi i m_0}$, and $F_l$ are Cartans of the flavor group. Now we claim that the elliptic genus of the strings in 5d is just given by setting $\epsilon_1=0$ in (\ref{eq:6dgenus}) which corresponds to the Nekrasov-Shatashvili limit \cite{Nekrasov:2009rc}. In particular it is defined by the operation
\begin{equation} \label{eq:res}
	Z^{\textrm{NS}}_r(Q_{\tau},\epsilon_2,m_l) = \widetilde{\textrm{Res}}_{2\pi i \epsilon_1=0} Z_r(Q_{\tau},\epsilon_1,2\epsilon_2,m_l),
\end{equation}
where we identify the Cartan generator of $SO(3)_{\mathbb{R}^3}$ with $J_2$ and the factor $2$ is introduced for conventional reasons. Now it is crucial to observe that equation (\ref{eq:res}) is not the ordinary residue operation. The ordinary residue does not transform covariantly under $SL(2,\mathbb{Z})$ if $Z_r$ has higher than single order poles in $\epsilon_1$ in its Laurent expansion. In order to take this into account we define the modified residue operation $\widetilde{\textrm{Res}}$ which gives rise to an $SL(2,\mathbb{Z})$ modular object. To this end we observe that the elliptic genus of $r$ strings, denoted by $Z_r$, has a representation in the following form:
\begin{equation}
	Z_r = \sum_{d=1}^r \varphi_d(\tau,\epsilon_1,\epsilon_2,m_l),
\end{equation}
where the functions $\varphi_d$ with $d=1, \ldots, r$ are meromorphic Jacobi forms with poles of order $d$ at $\epsilon_1=0$. Note that the above expansion is unique as all $\varphi_d$ have an order $2r$ pole at $\epsilon_1 = -\epsilon_2 = \epsilon = 0$ (unrefined limit) which reflects the fact that $r$ strings are probing $\mathbb{R}^4$. We can now define our residue to be given by
\begin{equation}
	\widetilde{\textrm{Res}}_{2\pi \epsilon_1=0} Z_r = \textrm{Res}_{2\pi i \epsilon_1=0} \varphi_1.
\end{equation}
This leads to a unique Jacobi form with elliptic parameters $\epsilon_2$ and $m_l$. In the next subsection we will elaborate on this claim by defining BPS generating functions.

\subsection{BPS degeneracies}
\label{sec:BPS}

In this section we want to define generating functions for BPS degeneracies which we identify with protected spin characters associated to the excitations of the BPS strings in 5d. To this end recall that the full partition function of the six-dimensional theory on $T^2 \times \mathbb{R}^4$ can be identified with the refined topological string partition function of the Calabi-Yau $X$:
\begin{equation}
	Z_{\textrm{top}} = 1 + \sum_{r=1}^{\infty} e^{-r \phi} Z_r(Q_{\tau},\epsilon_1,\epsilon_2,m_l).
\end{equation} 
To connect to strings in 5d we define the NS limit of the above partition function\footnote{For conventional reasons and to connect to the literature about Poincar\'e polynomials the dependence on $\epsilon_2$ has a factor of $2$ in the right-hand side of equation (\ref{eq:ResOp}).}:
\begin{equation} \label{eq:ResOp}
	Z^{\textrm{NS}}_{\textrm{top}}(Q_{\tau},\epsilon_2,m_l) = \widetilde{\textrm{Res}}_{2\pi i \epsilon_1 = 0} Z_{\textrm{top}}(Q_{\tau},\epsilon_1,2\epsilon_2,m_l),
\end{equation}
 On the other hand, we also know that BPS degeneracies are associated (up to multi-covering) with expansion coefficients of free energies of the topological string. These are defined by taking the logarithm of the partition function
\begin{equation}
	F_{\textrm{top}} = \log\left(Z^{\textrm{NS}}_{\textrm{top}}\right) = \sum_{r=1}^{\infty} e^{-r \phi} F_r(Q_{\tau},\epsilon_2,m_l),
\end{equation}
where the relation between the $F_r$ and $Z_r$ is given as follows
\begin{equation}
	F_r(Q_{\tau},\epsilon_2,m_l) = \sum_{r_1 + \cdots r_k = r} \frac{(-1)^k}{k} \prod_{i=1}^k Z^{\textrm{NS}}_{r_i}(Q_{\tau},\epsilon_2,m_l).
\end{equation}
This relation is reminiscent  to the relation between BPS invariants and virtual Poincar\'e polynomials as given in \cite{Joyce:2004tk,Manschot:2014cca}. In fact we take this relation seriously and define $F_r$ as generating functions for invariants $\bar{\Omega}$:
\begin{equation}
	F_r(Q_{\tau},\epsilon_2,m_l) = \sum_{n,i_l} \bar{\Omega}(\{r,n,i_l\},t,y) Q_{\tau}^n \prod_{l=1}^N Q_{m_l}^{i_l},
\end{equation}
where $t= e^{-2\pi i \epsilon_2}$, $Q_{m_l} = e^{2\pi i m_l}$ . Furthermore, $\{r,n,i_l\} \equiv \gamma$ denotes the charge vector associated to the BPS strings. The invariants $\bar \Omega$ can further be related to protected spin characters as in \cite{Manschot:2014cca} (see also \cite{Gopakumar:1998ii,Gopakumar:1998jq} for similar definitions in a different context):
\begin{equation}
	\Omega(\gamma,t,y) = \sum_{m | \gamma} \frac{\mu(m)}{m} \bar \Omega(\gamma/m,-(-t)^m,y^m),
\end{equation}
where $\mu(m)$ is the arithmetic M\"obius function. The function $\Omega(\gamma,t,y)$, known as the BPS degeneracy, is defined in \cite{Gaiotto:2010be}
\begin{equation} \label{eq:BPS}
	\Omega(\gamma,t,y) = \textrm{Tr}_{\textrm{BPS}} (-t)^{2J_2} y^{2J_I}.
\end{equation}
The indices $\bar \Omega$ and $\Omega$ are in fact subject to wall-crossing but we have chosen to neglect the dependence on the modulus as the topological string free energies compute these indices at a certain point in the moduli space which we shall clarify later. Furthermore, in cases where no exotics are present in the spectrum \cite{DelZotto:2014bga}, the index $\Omega(\gamma,-t,-t)$ is independent of $J_I$ and is related to the Poincar\'e polynomial of moduli spaces of sheaves on surfaces \cite{Hosono:2001gf} where $J_2$ is interpreted as the Cartan of the $SU(2)$ Lefschetz action on the cohomology. For us it turns out to be more convenient to define the following quantity\footnote{More generally we have $(t-t^{-1})\Omega(\gamma,t,y) = \sum_{m,n \in \mathbb{Z}}(-t)^{m+n-\textrm{dim}_{\mathbb{C}}\mathcal{M}} y^{m-n} h^{m,n}(\mathcal{M}(\gamma))$, where we have identified $2 J_2 = m + n - \textrm{dim}_{\mathbb{C}}(\mathcal{M})$ with the Cartan of the $SU(2)$ Lefschetz action on the Dolbeault cohomology and $2J_I = m - n$ with the "Hodge" $SU(2)$.}
\begin{equation}
	\Omega(\gamma,t) \equiv \Omega(\gamma,t,1) = \frac{P(\gamma,t)}{t-t^{-1}},
\end{equation}
with
\begin{equation}
	P(\gamma,t) = t^{-\textrm{dim}_{\mathbb{C}}\mathcal{M}(\gamma)} \sum_{l=0}^{2\textrm{dim}_{\mathbb{C}}\mathcal{M}(\gamma)} (-1)^l b_l(\mathcal{M}(\gamma))t^l.
\end{equation}
These Poincar\'e polynomials are associated to moduli spaces of sheaves on the surface $D$ as the M5 brane which gives rise to the strings in 5d is wrapping $T^2 \times D$ and when we take the size of $T^2$ to be small we obtain the partition function of twisted $\mathcal{N}=4$ $SU(r)$ Yang-Mills theory \cite{Vafa:1994tf} on $D$ as observed in \cite{Minahan:1998vr}. The associated Euler numbers are then equivalent to the instanton numbers of the twisted Yang-Mills theory \cite{Minahan:1998vr} which are obtained by taking the limit
\begin{equation}
	\Omega(\gamma) = \lim_{t\rightarrow 1} (t-t^{-1})\Omega(\gamma,t).
\end{equation}
The fact that the Nekrasov-Shatashvili limit of the refined topological string on the Calabi-Yau $X$ corresponds to generating functions for Poincar\'e polynomials of sheaves on $D$ was already observed for rank one sheaves and $D=\frac{1}{2}K3$ in \cite{Hosono:2001gf} and conjectured to hold for arbitrary rank for that case in \cite{Klemm:2012sx}. The elliptic genus of the 6d strings now provides a way to test this conjecture for other $X$ and $D$ and for all ranks $r$ since its knowledge is equivalent to the all genus result of the refined topological string. We shall test our claim by using the elliptic genus of M-strings to obtain the elliptic genus of strings of the 5d $\mathcal{N}=1^*$ $SU(2)$ theory in the next section. 

\section{Strings of 5d $\mathcal{N}=1^*$ $SU(2)$ gauge theory}

To learn about strings of 5d $\mathcal{N}=1^*$ gauge theory we need to study strings of the 6d $(2,0)$ theory, namely M-strings. Their elliptic genus was computed in \cite{Haghighat:2013gba} by using the topological vertex formalism and by a 2d quiver gauge theory in \cite{Haghighat:2013tka}. The result for $r$ M-strings can be written in a compact form as follows
\begin{equation} \label{eq:MstrRes}
	Z_r(\tau,\epsilon_1,\epsilon_2, m) = \sum_{|\nu|=r} \prod_{(i,j)\in \nu} \frac{\theta_1(\tau;z_{ij})\theta_1(\tau;v_{ij})}{\theta_1(\tau;w_{ij})\theta_1(\tau;u_{ij})},
\end{equation}
where $\nu$ is a Young tableau and $|\nu|$ denotes the number of boxes in $\nu$. Furthermore, $(i,j) \in \nu$ specifies a box in the $i$'th row and $j$'th column and the elliptic parameters as functions of $y = Q_m = e^{2\pi i m}$, $q=e^{2\pi i \epsilon_1}$ and $t = e^{-2\pi i \epsilon_2}$ are given by
\begin{equation}
\begin{array}{ll}
	e^{2\pi z_{ij}} = Q_m^{-1} q^{\nu_i - j + 1/2}t^{-i+1/2}, & e^{2\pi i v_{ij}} = Q_m^{-1} t^{i-1/2}q^{-\nu_i + j-1/2}, \nonumber \\
	e^{2\pi i w_{ij}} = q^{\nu_i -j +1} t^{\nu_j^t - i},  & e^{2\pi u_{ij}} = q^{\nu_i -j} t^{\nu_j^t - i +1},
	\end{array}
\end{equation}
and the theta function is defined as
\begin{equation}
	\theta_1(\tau;z) = i Q_{\tau}^{1/8} e^{\pi i z} \prod_{k=1}^{\infty}(1-Q_{\tau}^k)(1-Q_{\tau}^k e^{2\pi i z}) (1-Q_{\tau}^{k-1}e^{-2\pi i z}).
\end{equation}
In order to obtain the elliptic genus of the 5d strings we have to apply the residue operation of (\ref{eq:ResOp}) to (\ref{eq:MstrRes}). It can be easily seen that the result is the following
\begin{equation} \label{eq:5dres}
	Z^{\textrm{NS}}_r(\tau,\epsilon_2,m) = - i \frac{\prod_{k=-r}^{r-1}\theta_1(\tau;-m+(2k+1)\epsilon_2)}{\eta(\tau)^3 \theta_1(\tau;2r\epsilon_2)\prod_{k=1}^{r-1}\theta_1(\tau;2k\epsilon_2)^2},
\end{equation}
where the only contribution in (\ref{eq:res}) comes from fully anti-symmetric Young tableaux and $\epsilon_2$ got rescaled by a factor of $2$ in accord with the definition (\ref{eq:ResOp}). 
Now one can convince oneself that this result agrees for $m=0$ exactly with the result obtained in \cite{Mozgovoy:2013zqx} for the generating function of rank $r$ sheaves on a rationally ruled surface with an elliptic curve as base\footnote{The expression (\ref{eq:5dres}) for the generating function of rank $r$ sheaves on $\mathbb{P}^1 \times T^2$ was already conjectured in \cite{Manschot:2011ym} (Conjecture 4.3).}. This surface is the surface $D$ described in section \ref{sec:6d5d} for the geometric engineering of the 5d $\mathcal{N}=1^*$ $SU(2)$ theory. As it is just the direct product $D \cong \mathbb{P}^1 \times E$ one can view the $\mathbb{P}^1$ as the rational fiber and $E$ as the elliptic curve base. Henceforth we will denote the $\mathbb{P}^1$ by $f$. Then the exact result in \cite{Mozgovoy:2013zqx} is that (\ref{eq:5dres}) is the generating function for rank $r$ semi-stable sheaves on $D$ with K\"ahler class $J=f$ and $c_1 \cdot f = 0$ mod $r$. In order to derive (\ref{eq:5dres}) from the expression given in \cite{Mozgovoy:2013zqx} one has to use that the motivic zeta function of an elliptic curve is given by:
\begin{equation}
	Z_E(\tau,t) = \frac{(1-t Q_{\tau})(1-t Q_{\tau})}{(1-Q_{\tau})(1-t^2 Q_{\tau})}.
\end{equation}
The choice $J=f$ is at the boundary of the K\"ahler cone. In order to connect to the weak coupling chamber of the four-dimensional gauge theory which arises by circle compactification from the five-dimensional one, one has to move slightly away from the boundary as argued in \cite{Haghighat:2012bm} (see also \cite{Chuang:2013wt}). This is done by choosing 
\begin{equation} \label{eq:infwall}
	J = f + \delta E,
\end{equation}
with $\delta \ll 1$.

\subsection{Magnetic charge $r=1$}

For $r=1$ we obtain the following \textrm{NS}-limit of the M-string elliptic genus:
\begin{equation} \label{eq:onestring}
	Z_1^{\textrm{NS}}(\tau,\epsilon_2,m) = \frac{1}{\eta(\tau) \theta_1(\tau;-2\epsilon_2)}\frac{\theta_1(\tau;-m+\epsilon_2)\theta_1(\tau;-m-\epsilon_2)}{\eta(\tau)^2}.
\end{equation}
This result has a nice interpretation in terms of the moduli space of magnetic monopoles of magnetic charge $r=1$. Their bosonic moduli space is given by $\mathbb{R}^3 \times S^1$ which explains the factor 
\begin{equation}
	\frac{1}{\eta(\tau)\theta_1(\tau;-2\epsilon_2)},
\end{equation}
with $\eta(\tau)$ corresponding to the boson parametrizing $S^1$ and $\theta_1(\tau;-2\epsilon_2)$ corresponding to $\mathbb{R}^3$ on which $U(1)_{\epsilon_2}$ acts equivariantly. Moreover, the second factor, namely
\begin{equation}
	\frac{\theta_1(\tau;-m+\epsilon_2)\theta_1(\tau;-m-\epsilon_2)}{\eta(\tau)^2},
\end{equation}
corresponds to the four fermionic zero-modes which are present due to the adjoint hypermultiplet of the 5d $\mathcal{N}=1^*$ theory \cite{Callias:1977kg}. We can also expand expression (\ref{eq:onestring}) in powers of $Q_{\tau}$:
\begin{eqnarray} \label{eq:onestringexp}
	Z_1^{\textrm{NS}}(\tau,\epsilon_2,m) & = & \frac{(Q_m-t)(Q_m t - 1)}{Q_m t (t-t^{-1})} - \frac{(Q_m-t)^2(Q_m t - 1)^2 (1+t^2)}{Q_m^2 t^3(t-t^{-1})} Q_{\tau} \nonumber \\
	~ & ~ & + \frac{(Q_m-t)^2(Q_m t-1)^2(t^3+Q_m^2 t^3 - Q_m (1+t^2)^3)}{Q_m t^5(t-t^{-1})} Q_{\tau}^2 + \mathcal{O}(Q_{\tau}^3). \nonumber \\
\end{eqnarray}
As $\tau = i \frac{R_1}{R_2}$ where $R_2$ is the radius of the circle compactifying the five-dimensional theory to four dimensions, we see that $Q_{\tau} = 0$ corresponds to the four-dimensional limit while higher powers of $Q_{\tau}$ correspond to higher momentum along $S^1_{R_1}$. When we set the mass $m$ of the adjoint hypermultiplet to zero, i.e. when we go to the limit of the maximal SYM theory in five dimensions, we can extract the Betti numbers shown in Table \ref{tb:betti1}.
\begin{table}[h!] 
\begin{center}
\begin{tabular}{|c|c|c|c|c|c|c|c|c|c|c|c|}
    \hline
    $Q_{\tau}^n$ & $b_0$ & $b_1$ & $b_2$ & $b_3$ & $b_4$ & $b_5$ & $b_6$ & $b_7$ & $b_8$ & $b_9$ & $b_{10}$\\
    \hline
    $n=0$ & $1$ & $2$ & $1$ & ~     & ~    & ~ & ~ & ~ & ~ & ~ & ~ \\
    \hline
    $n=1$ & $1$ & $4$ & $7$ & $8$ & $7$ & $4$ & $1$ & ~ & ~ & ~ & ~ \\
    \hline
    $n=2$ & $1$ & $4$ & $9$ & $18$ & $30$ & $36$ & $30$ & $18$ & $9$ & $4$ & $1$ \\
    \hline 
\end{tabular} 
\end{center} 
\caption{Betti numbers for $r=1$ and $m=0$.}
\label{tb:betti1}
\end{table}

One can see that the Betti numbers in the above table give zero Euler numbers. This is to be expected as the string of maximal SYM in 5d has $(4,4)$ world-sheet supersymmetry instead of $(0,4)$. Therefore, the left-moving side is also supersymmetric and leads to a perfect cancellation between fermions and bosons. The mass-deformation breaks the left-moving supersymmetry and can be thought of as introducing a grading in the cohomology ring of the moduli space of the string given by powers of $Q_m$. In order to define an elliptic genus for the $(4,4)$ string reference \cite{Kim:2011mv} proposed to set $Q_m = -1$. However, recalling that the mass $m$ couples to the R-symmetry current $J_I$, one can extract from (\ref{eq:onestringexp}) the following more refined information (see also page $4$ of \cite{Diaconescu:2007bf} for an explanation of the relation between Hodge numbers, spin and R-symmetry):

\begin{table}[h!]
\begin{center}
\begin{tabular}{|c|c|c|c|}
    \hline
    $k$ & $b_0^k$ & $b_1^k$ & $b_2^k$ \\
    \hline
    $0$ & $1$ & ~ & $1$ \\
    \hline
    $1$ & ~ & 1 & ~  \\
    \hline 
\end{tabular}
\end{center}
\caption{Refined Betti numbers 
$b^k_i =h^{\frac{i+k}{2},\frac{i-k}{2}}$ for $r=1$ and $Q_{\tau}^0$.}
\end{table}

\begin{table}[h!]
\begin{center}
\begin{tabular}{|c|c|c|c|c|c|c|c|}
    \hline
    $k$ & $b_0^k$ & $b_1^k$ & $b_2^k$ & $b_3^k$ & $b_4^k$ & $b_5^k$ & $b_6^k$\\
    \hline
    $0$ & $1$ & ~ & $5$ & ~ & $5$ & ~ & $1$ \\
    \hline
    $1$ & ~ & $2$ & ~ & $4$ & ~ & $2$ & ~  \\
    \hline 
    $2$ & ~ & ~ & $1$ & ~ & $1$ & ~ & ~  \\
    \hline
\end{tabular}
\end{center}
\caption{Refined Betti numbers 
$b^k_i =h^{\frac{i+k}{2},\frac{i-k}{2}}$ for $r=1$ and $Q_{\tau}^1$.}
\end{table}

\begin{table}[h!]
\begin{center}
\begin{tabular}{|c|c|c|c|c|c|c|c|c|c|c|c|}
    \hline
    $k$ & $b_0^k$ & $b_1^k$ & $b_2^k$ & $b_3^k$ & $b_4^k$ & $b_5^k$ & $b_6^k$ & $b_7^k$ & $b_8^k$ & $b_{9}^k$ & $b_{10}^k$ \\
    \hline
    $0$ & $1$ & ~ & $7$ & ~ & $20$ & ~ & $20$ & ~ & $7$ & ~ & $1$\\
    \hline
    $1$ & ~ & $2$ & ~ & $9$ & ~ & $17$ & ~ & $9$ & ~ & $2$ & ~  \\
    \hline 
    $2$ & ~ & ~ &  $1$ & ~ & $5$ & ~ & $5$ & ~ & $1$ & ~  & ~ \\
    \hline
    $3$ & ~ & ~ & ~ & ~   & ~      & $1$ & ~ & ~ & ~ & ~ & ~ \\
    \hline
\end{tabular}
\end{center}
\caption{Refined Betti numbers 
$b^k_i =h^{\frac{i+k}{2},\frac{i-k}{2}}$ for $r=1$ and $Q_{\tau}^2$.}
\end{table}

\subsection{Magnetic charge $r=2$}

The elliptic genus for two strings takes the following form:
\begin{equation}
	Z_2^{\textrm{NS}}(\tau,\epsilon_2,m) = - i \frac{\theta_1(\tau;-m+3 \epsilon_2)\theta_1(\tau;-m+ \epsilon_2)\theta_1(\tau;-m-\epsilon_2)\theta_1(\tau;-m-3 \epsilon_2)}{\eta(\tau)^3 \theta_1(\tau;-2\epsilon_2)^2\theta_1(\tau;-4 \epsilon_2)}.
\end{equation}
Following the rules of section \ref{sec:BPS} we can extract from this the following BPS generating function:
\begin{equation} \label{eq:2monA}
	Z_2^{\textrm{NS}}(\tau,\epsilon_2,m) - \frac{1}{2} Z_1^{\textrm{NS}}(\tau,\epsilon_2,m)^2 + \frac{1}{2} Z_1^{\textrm{NS}}(2\tau,1/2+2\epsilon_2,2m),
\end{equation}
where the last term is added to subtract multi-covering contributions. However, as argued in \cite{Haghighat:2012bm}, one has to add to the above expression the following term
\begin{equation} \label{eq:2monB}
	\frac{1}{1-t^4} Z_1^{\textrm{NS}}(\tau,\epsilon_2,m)^2,
\end{equation}
which is due to moving infinitesimally away from the boundary of the K\"ahler cone (\ref{eq:infwall}), in order to arrive in the weak coupling chamber of the field theory. Expanding the sum of (\ref{eq:2monA}) and (\ref{eq:2monB}) in powers of $Q_{\tau}$ we can again obtain Betti numbers. One notices that the expansion starts with the first power of $Q_{\tau}$ which is consistent with fact that the four-dimensional limit of our five-dimensional field theory is the $SU(2)$ $\mathcal{N}=2^*$ theory and is known to not have any bound states with magnetic charge greater than one. At first order in $Q_{\tau}$ we then obtain the Betti numbers shown in table \ref{tab:r2numbers}.
\begin{table}[h]
\begin{center}
\begin{tabular}{|c|c|c|c|c|c|c|c|c|c|c|c|}
    \hline
    $Q_{\tau}^n$ & $b_0$ & $b_1$ & $b_2$ & $b_3$ & $b_4$ & $b_5$ & $b_6$ & $b_7$ & $b_8$ & $b_9$ & $b_{10}$\\
    \hline
    $n=1$ & $1$ & $4$ & $7$ & $8$ & $8$ & $8$ & $8$ & $8$ & $7$ & $4$ & $1$ \\
    \hline
\end{tabular}
\end{center} 
\caption{Betti numbers for $r=1$ and $m=0$.}
\label{tab:r2numbers}
\end{table}
Again one notices that the corresponding Euler number is zero. Refining the information by taking into account the mass-deformation one arrives at the numbers shown in table \ref{tab:r2refnumbers}.
\begin{table}[h!]
\begin{center}
\begin{tabular}{|c|c|c|c|c|c|c|c|c|c|c|c|}
    \hline
    $k$ & $b_0^k$ & $b_1^k$ & $b_2^k$ & $b_3^k$ & $b_4^k$ & $b_5^k$ & $b_6^k$ & $b_7^k$ & $b_8^k$ & $b_9^k$ & $b_{10}^k$\\
    \hline
    $0$ & $1$ & ~ & $5$ & ~ & $6$ & ~ & $6$ & ~ & $5$ & ~ & $1$ \\
    \hline
    $1$ & ~ & $2$ & ~ & $4$ & ~ & $4$ & ~ & $4$ & ~ & $2$ & ~  \\
    \hline 
    $2$ & ~ & ~ & $1$ & ~ & $1$ & ~ &  $1$ & ~ & $1$ & ~ & ~ \\
    \hline
\end{tabular}
\end{center}
\caption{Refined Betti numbers 
$b^k_i =h^{\frac{i+k}{2},\frac{i-k}{2}}$ for $r=2$ and $Q_{\tau}^1$.}
\label{tab:r2refnumbers}
\end{table}

\section{Concluding thoughts}

In this paper we have presented a proposal for how to obtain elliptic genera of strings in five-dimensional field theory from strings in a six-dimensional parent theory. The main restrictions we imposed on the elliptic genus are $SL(2,\mathbb{Z})$ modularity and the connection to sheaf counting on ruled surfaces. Our prescription naturally fulfils both requirements and is also consistent with knowledge about moduli spaces and bound states of magnetic monopoles. Moreover, an interesting advantage of the definition (\ref{eq:ResOp}) is that it easily connects to sheaf counting for more general ruled surfaces of the form $\mathbb{P}^1 \times \Sigma_g$ where $\Sigma_g$ is a genus $g$ Riemann surface \cite{AhsanAmerBabak}.

However, the question remains whether it is unique, that is, whether there is any other way to obtain an elliptic genus which also fulfils the above mentioned requirements? In fact, it turns out that there is another prescription as explored in reference \cite{Hohenegger:2015cba}. In \cite{Hohenegger:2015cba} the authors define the elliptic genus of 5d strings by first computing the topological string free energy on the relevant elliptic Calabi-Yau manifold
\begin{equation}
	\mathcal{F} = \log\left(Z_{\textrm{top}}\right) = \sum_{r=1}^{\infty} e^{-r \phi} \mathcal{F}_r(Q_{\tau},\epsilon_1,\epsilon_2,m_l),
\end{equation}
and then take the Nekrasov-Shatashvilli limit to obtain the elliptic genus\footnote{We again rescale $\epsilon_2$ by a factor of $2$ to adjust to the definition of Poincar\'e Polynomials.}:
\begin{equation} \label{eq:amer}
	\mathcal{F}^{\textrm{NS}}_r = \textrm{Res}_{2\pi i \epsilon_1=0} \mathcal{F}_r(Q_{\tau},2\epsilon_2,m_l).
\end{equation}
This definition equally gives an $SL(2,\mathbb{Z})$ invariant result as $\mathcal{F}_r$ has only simple poles in $\epsilon_1$. Note though, that the result obtained is different from the definition presented in this paper as the two operations ``taking the logarithm" and ``taking the NS-limit" do not commute. Also, it is not clear how (\ref{eq:amer}) connects to sheaf counting. Let us compare the results of the two prescriptions for one and two strings. For a single string they both agree as (\ref{eq:amer}) also gives the result (\ref{eq:onestring}). For two strings, however, we find\footnote{We suppress the $\tau$-dependence in the following.}
\begin{eqnarray}
	\mathcal{F}^{\textrm{NS}}_2 & = & Z^{\textrm{NS}}_2 -\frac{\theta_1(-m-\epsilon_2)\theta_1'(-m-\epsilon_2)\theta_1(-m+\epsilon_2)^2}{2\eta^6 \theta_1(-2\epsilon_2)^2} \nonumber \\
	~ & ~ & + \frac{\theta_1(-m-\epsilon_2)^2\theta_1(-m+\epsilon_2)\theta_1'(-m+\epsilon_2)}{2\eta^6 \theta_1(-2 \epsilon_2)^2} - \frac{\theta_1(-m-\epsilon_2)^2\theta_1(-m+\epsilon_2)^2\theta_1'(-2\epsilon_2)}{2\eta^6 \theta_1(-2\epsilon_2)^3} \nonumber \\
\end{eqnarray}
We see that the first term agrees with our prescription but the second term is different from $-\frac{1}{2} Z^{\textrm{NS}}_1(\tau,\epsilon_2,m)^2$ with the difference being
\begin{eqnarray}
	~ & ~ & -\frac{1}{2}\frac{\theta_1(-m-\epsilon_2)\theta_1(-m+\epsilon_2)}{\eta^6 \theta_1(-2\epsilon_2)^2} \times \left(\theta_1(-m-\epsilon_2)\theta_1(-m+\epsilon_2) + \theta_1'(-m-\epsilon_2)\theta_1(-m+\epsilon_2)\right. \nonumber \\
	~ & ~ &\left. - \theta_1'(-m+\epsilon_2)\theta_1(-m-\epsilon_2)+ \frac{\theta_1'(-2\epsilon_2)}{\theta_1(-2\epsilon_2)}\right). \nonumber \\
\end{eqnarray}
It would be very interesting to see whether this difference is due to wall-crossing in the moduli space of sheaves by moving from the edge to the interior of the K\"ahler cone of the ruled surface. In such a case, the prescription (\ref{eq:amer}) would equally connect to sheaf counting and the two different definitions for elliptic genera would be naturally connected.

\section*{Acknowledgements}
We would like to thank Clay Cordova, Michele Del Zotto, Amer Iqbal, Jan Manschot and Cumrun Vafa for valuable discussions. The work of B.H. is supported by the NSF grant DMS-1159412.

\end{document}

8